\documentstyle[11pt]{article}

\begin{document}
\vspace{.5cm}
\centerline{\bf\large{\bf{Supersymmetric  Harry Dym Type Equations }}}
\vspace{1in}
\centerline{\bf\large Q.\ P. \ Liu}
\smallskip
\centerline{CCAST(World Laboratory),}
\centerline{P.O.Box 8730, Beijing 100080, China}
\smallskip
\centerline{\bf\large and}
\smallskip

\centerline{Department of Mathematics,}
\centerline{Beijing Graduate School(\# 65),}
\centerline{China University of Mining and Technology,}
\centerline{Beijing 100083, China\footnote{mailing address}}
\newcommand{\be}{\begin{equation}}
\newcommand{\ee}{\end{equation}}
\newcommand{\ba}{\begin{array}}
\newcommand{\ea}{\end{array}}
\def\p{\partial}
\def\alf{\alpha}
\def\bi{\beta}
\def\es{\epsilon}
\def\la{\lambda}
\def\fr{\frac}
%\maketitle
\vspace{0.5in}\begin{center}
\begin{minipage}{5in}
{\bf ABSTRACT}\hspace{.2in} A supersymmetric version is proposed for
the well known Harry Dym system. A general class super Lax operator
which leads to consistent equations is considered.
\par
\end{minipage}
\end{center}
\vspace{.5in}
\vfill\eject
During last ten years or so, super extensions of integrable models have
been a subject arrested much attention. The consequence of such study is
that a number of the well known integrable systems are embedded in the
context of super systems. In particular, we mention here the super
Sine-Gordon[1], Korteweg-de Vries(KdV)[2,3,4] and nonlinear Schrodinger
equations[5] and super Kadomtsev-Petviashvili hierarchy[2],
etc..(see [6] for more references).
\par

We note that two type super extensions for a given integrable system may
exist, that is, supersymmetric and fermionic extensions. In KdV case, this
corresponds Manin-Radul' version[2] and Kupershmidt's version[4]
respectively. Apart from KdV system, Harry Dym(HD) equation is also
well known. Very recently, it is found that HD equation is not just a
mathematically interested model and it possesses physical applications[7].
Thus, it is interesting to construct a super analogy of HD equation.
In this regard, a fermionic HD model is known from Kupershmidt's work[8]
while a generic supersymmetric HD (sHD) system is still lacking to the best
of my knowledge. The aim of this Note is to propose such a model.

For convenience, we fix our notations at the very beginning: denoting
even variables by Latin letters and odd variables by Greek letters;
index $_{\geq r}$ of a operator always means the projection to the
part of order greater than $D^r$(including the term $D^r$ itself).

Let us first recall some basic facts of HD equation. The equation
reads

\be
w_t=\fr{1}{4}w^3w_{xxx}
\ee
or
\be
u_t=\fr{1}{4}(u^{\fr{3}{2}}u_{xxx}-\fr{3}{2}u^{\fr{1}{2}}u_xu_{xx}
+\fr{3}{4}u^{-\fr{1}{2}}u_x^{3})
\ee
The link between them is $u=w^2$.

It is known that HD equation(2) has the following Lax operator
\be
L_{HD}=u\p^2
\ee
and the Lax equation is
\be
L_{HD_{t}}=[P, L_{HD}]
\ee
where $P=(L^{\fr{3}{2}})_{{\geq}4}$.

\par
Our candidate for the supersymmetric Harry Dym(sHD) equation is:
\be
\begin{tabular}{ll}
&$
u_t=(\fr{1}{4}u^{\fr{3}{2}}u_{xxx}-\fr{3}{8}u^{\fr{1}{2}}u_xu_{xx}+
\fr{3}{16}u^{-\fr{1}{2}}u_{x}^{3})+\fr{3}{8}u^{\fr{1}{2}}\alf (Du_{xx})-
\fr{3}{4}u^{\fr{1}{2}}\alf \alf_{xx} -$\\[2mm]
&$~~~~~~\fr{3}{32}u^{-\fr{3}{2}}u_{x}^{2}\alf(Du)+
\fr{3}{16}u^{-\fr{1}{2}}u_{xx}\alf(Du)+
\fr{3}{8}u^{-\fr{1}{2}}u_x\alf_x(Du)-
\fr{3}{8}u^{-\fr{1}{2}}u_x\alf(Du_x)-$\\[2mm]
&$~~~~~~
\fr{3}{8}u^{\fr{1}{2}}\alf_{xx}(Du)+
\fr{3}{16}u^{-\fr{3}{2}}u_x\alf(D\alf)(Du)-
\fr{3}{8}u^{-\fr{1}{2}}\alf (D\alf_x)(Du)+
\fr{3}{4}u^{-\fr{1}{2}}\alf \alf_x u_x.$\\[2mm]

&$\alf_t=\fr{1}{4}u^{\fr{3}{2}}\alf_{xxx}+
\fr{3}{8}u^{\fr{1}{2}}\alf(D\alf_{xx})+
\fr{3}{16}u^{-\fr{1}{2}}u_{x}^{2}\alf_x-
\fr{3}{8}u^{\fr{1}{2}}u_{xx}\alf_x-
\fr{9}{32}u^{-\fr{3}{2}}u_{x}^{2}\alf(D\alf)+$\\[2mm]
&$~~~~~~\fr{3}{16}u^{-\fr{1}{2}}u_{xx}\alf(D\alf)+
\fr{3}{8}u^{-\fr{1}{2}}u_x\alf_x(D\alf)-
\fr{3}{8}u^{\fr{1}{2}}\alf_{xx}(D\alf)+
\fr{3}{16}u^{-\fr{3}{2}}u_x\alf(Du)\alf_x-$\\[2mm]
&$~~~~~~\fr{3}{8}u^{-\fr{1}{2}}\alf(Du_x)\alf_x
+\fr{3}{16}u^{-\fr{3}{2}}u_x\alf(D\alf)^2+
\fr{3}{16}u^{-\fr{3}{2}}\alf(Du)\alf_x(D\alf)-
\fr{3}{8}u^{-\fr{1}{2}}\alf(D\alf)(D\alf_x)$
\end{tabular}
\ee
Where $D=\theta \p +\p_{\theta}$, u is a super even variable
and $\alf$ is a super odd variable. Since the system is
formulated in superderivartive and super fields, the
supersymmetry is manifest. If we set the odd field
variable $\alf$ to zero, we get the HD equation(2),
thus the system (5) deserves the name of sHD.

The system(5) has the following Lax representation:
\be
L_t=[P,L]
\ee
where $L=u\p^2+\alf \p D$, and $P=(L^{\fr{3}{2}})_{{\geq}3}=
u^{\fr{3}{2}}\p^3+\fr{3}{2}u^{\fr{1}{2}}\alf\p^2D+
(\fr{3}{4}u^{\fr{1}{2}}u_x+\fr{3}{8}u^{-\fr{1}{2}}\alf(Du))\p^2+
(\fr{3}{4}u^{\fr{1}{2}}\alf_x+\fr{3}{8}u^{-\fr{1}{2}}\alf(D\alf))\p D.$.

Noticing the sHD system can be reformulated in Lax form, we see that
this kind of representation is nonstandard in Kupershmidt
sense[9](see also Kiso[10]). A detailed presentation of
nonstandard Lax representation can be found in[11].
This remark suggests us to consider more general operator:
\be
L=u\p^2+\alf \p D+v\p+\bi D+w
\ee
Taking L as a Lax operator, we may construct integrable
systems by means of fractional power method. It is not
difficult to verify that the following four cases occur:

{\em Case 1}:
\be
L_t=[(L^{\fr{3}{2}})_{{\geq}0},L]
\ee
The  standard argument shows that the system is consistent:
since $[(L^{\fr{3}{2}})_{{\geq}0},L]=-[(L^{\fr{3}{2}})_{<0},L]$,
the right hand side of (8) is a diffenertial operator of form
$A\p +\gamma D+B$. Thus, we may set $u=1, \alf=0$. However, this
implies $v=0$ and we end with Manin-Radul case[2,3].

{\em Case 2}:
\be
L_t=[(L^{\fr{3}{2}})_{{\geq}1},L]
\ee
The same argument leads to the conclusion: we may set $u=1,
\alf=0$. Thus, we here have a system of three equations. It
is easy to see that we may further set $w=0$. This last case
was noticed by Inami and Kanno[6].

{\em Case 3}:
\be
L_t=[(L^{\fr{3}{2}})_{{\geq}2},L]
\ee
The general case will lead to a system involved five fields.
However, we may have a subsystem which only have three fields.

{\em Case 4}:
\be
L_t=[(L^{\fr{3}{2}})_{{\geq}3},L]
\ee
As above, this system involved all five fields. A reduction
gives us the sHD system(5).

\bigskip
{\em Remarks}.

(1). In the pure bosonic case, we only have three cases which
corresponds to KdV, MKdV and Harry-Dym systems respectively[10,11].

(2). A simple calculation shows that the next one $ L_t=
[(L^{\fr{3}{2}})_{{\geq }4},L]$ will not lead to any
consistent system. Thus, we here have only four cases.

\bigskip

For a general even order diffenertial  operator
\be
L=\sum_{i=0}^{n}u_iD^{2i}+\sum_{i=1}^{n}\alf_iD^{2i-1}
\ee
we may consider the following Lax equation
\be
L_{t}=[(L^{\fr{k}{n}})_{{\geq} r},L], \quad r=0, 1, 2, 3
\ee
The $r=0, 1$ cases are considered in [12] and [6]
respectively. It is pointed out that one may set the
fields $u_n=\alf_n=u_{n-1}=0$ to zero in the case $r=0$ and $u_n=\alf_n
=u_0=0$ in the case $r=1$. When $r=2$ and $r=3$, all the field
variables can be taken as dynamical variables. However, the following
reductions or restrictions are feasible: $u_0=0, \alf_1=0$ for $r=2$
case and $u_1=u_0=0, \alf_1=0$ for the case $r=3$.

The systems (13) are integrable in the sense that they
consist of commuting flows. The proof of this statement
is not difficult: the first two cases  are proved in the
cited references. The proof for last two cases follows
from the standard argument.

We conclude this Note with the remarks in order:

(1). We see that for the operator(12), we have four cases mentioned above.
This phenomenon is based on the following algebraic decompositions:
\be
g=\{\sum_{i} u_i D^i \}=g_{{\geq} r} \oplus g_{< r}, \quad r=0,1,2,3.
\ee

(2). It would be interesting to study the Hamiltonian
structures of our system sHD. We know that HD equation
is not only Hamiltonian but bi-Hamiltonian. We suspect
that it is also the case for the sHD.

(3). It is proved that the case $r=0$ and case $r=1$
are gauge related each other[6]. It is important to
study the relationship between $r=1$ case and the $r=2$
case and $r=3$ case.

(4). We may construct the flows in terms of Sato's
approach. The candidate of pseudo-diffenertial operator
is $L=u_0\p+\alf_0 D+ ...$.

(5). In a recent paper[14], Darboux transformations
for sKdV are constructed. The same consideration will
be interesting for general cases.
\bigskip

{\bf{Acknowledgement}}

%\end{document}
%\bye

I should like to thank Boris Kupershmidt for
communications on sHD. It is my pleasure to thank the referee for
the useful comments. This work is supported
by National Natural Science Foundation of China. \par
\bigskip
{\bf References}

\begin{flushleft}
\small
[1] Chainchian M and Kulish P P 1978 {\em Phys.Lett.} {\bf 78B} 413.\par
[2] Manin Yu and Rudal A 1985 {\em Commun. Math. Phys.} {\bf 98} 65.

[3] Mathieu P 1988 {\em J. Math. Phys.} {\bf 28} 2499.

[4] Kupershmidt B A 1984 {\em Phys. Lett.} {\bf 102A} 213.

[5] Roelofs G H M and Kersten P H M 1992 {\em J. Math. Phys.} {\bf 33} 2185.

[6] Inami T and Kanno H 1992 {\em Inter. J. Mod. Phys.}
{\bf 7} (Suppl 1A) 419;  Morois C and Pizzocchero L 1994
{\em J. Math. Phys.} {\bf 35} 2397.

[7] Kadanoff L P 1990 {\em Phys. Rev. Lett.} {\bf 65} 2986.

[8] Kupershmidt B A 1987 {\em Elements of Superintegrable Systems}, D.
Reidel, Dordrecht.

[9] Kupershmidt B A 1985 {\em Commun. Math. Phys.} {\bf 99} 51.

[10] Kiso K 1990 {\em Progr. Theor. Phys. } {\bf 83} 1108.

[11] Konopelchenko B G and Oevel W 1993 {\em Publ. RIMS, Kyoto
University} {\bf 29} 1.

[12] Figueroa-O'Farill J  ~Ramos E and  Mas J 1991 {\em Rev.
Math. Phys.} {\bf 3} 479; Oevel W and Popowicz Z 1991 {\em
Commun. Math. Phys.} {\bf 139} 441.\par
[13] Oevel W and Rogers C 1993 {\em Rev. Math. Phys.}
{\bf 5} 299.\par

[14] Liu Q P 1994 Darboux Transformation for
Supersymmetric  Korteweg-de Vries Equations,
preprint, ASITP-94-41, {\em Lett. Math. Phys.} (in press).
\end{flushleft}
\end{document}